\documentclass{interact}

\usepackage[utf8]{inputenc}
\usepackage[english]{babel}
\usepackage{amsmath} 
\usepackage{amsfonts} 
\usepackage{amssymb}
\usepackage{graphicx} 
\usepackage{xcolor}
\usepackage{epstopdf}
\usepackage{subfigure}

\newcommand{\mr}{\mathbf{r}}

\newcommand{\bra}[1]{\ensuremath{\langle #1 \vert}}
\newcommand{\ket}[1]{\ensuremath{\vert #1 \rangle}}
\renewcommand{\H}{\ensuremath{\text{H}}}

\newcommand{\la}{\ensuremath{\lambda}}
\newcommand{\md}{\ensuremath{\text{md}}}
\newcommand{\lr}{\ensuremath{\text{lr}}}
\newcommand{\sr}{\ensuremath{\text{sr}}}
\newcommand{\w}{w_{\text{ee}}}
\DeclareMathOperator{\erf}{erf}

\newcommand{\Hxc}{\text{Hxc}}
\newcommand{\Hx}{\text{Hx}}
\newcommand{\ee}{\text{ee}}
\renewcommand{\x}{\text{x}}
\newcommand{\tc}{\text{c}}

\newcommand{\onlinecite}{\cite}

\usepackage[numbers,sort&compress,merge]{natbib}
\bibpunct[, ]{[}{]}{,}{n}{,}{,}
\usepackage{hyperref}

\begin{document}
\title{Excitation energies from G\"orling--Levy perturbation theory along the
  range-separated adiabatic connection}

\author{
  Elisa Rebolini$^{1}$$^{\ast}$,
  \thanks{$^\ast$Email: rebolini@ill.fr \vspace{6pt}}
  \thanks{$^\dagger$Email: julien.toulouse@upmc.fr \vspace{6pt}}
  Andrew M. Teale$^{3,4,5}$,
  Trygve Helgaker$^{4,5}$, 
  Andreas Savin$^{2}$,
  and Julien Toulouse$^{2}$$^{\dagger}$,\\\vspace{6pt}
  $^1${\em{Institut Max von Laue - Paul Langevin,
      71, avenue des Martyrs, F-38042, Grenoble, France}}\\
  $^2${\em{Laboratoire de Chimie Th\'eorique, Universit\'e Pierre et Marie
      Curie, CNRS, Sorbonne Universit\'es, F-75005 Paris, France}}\\
  $^3${\em{School of Chemistry, University of
      Nottingham, University Park, Nottingham NG7 2RD, United Kingdom}}\\
  $^4${\em{Hylleraas Centre for Quantum Molecular Sciences, Department of Chemistry,
   University of Oslo, P.O. Box 1033 Blindern, N-0315 Oslo, Norway}}\\
  $^5${\em{Centre for Advanced Study at the Norwegian Academy of Science and Letters.
  Drammensveien 78, N-0271 Oslo, Norway}}
}

\maketitle

\begin{abstract}
A G\"orling--Levy (GL)-based perturbation theory along the range-separated adiabatic connection is assessed for the calculation of electronic excitation energies. In comparison with the Rayleigh--Schr\"odinger (RS)-based perturbation theory introduced in a previous work [E. Rebolini, J. Toulouse, A. M. Teale, T. Helgaker, A. Savin, Mol. Phys. {\bf 113}, 1740 (2015)], this GL-based perturbation theory keeps the ground-state density constant at each order and thus gives the correct ionization energy at each order. Excitation energies up to first order in the perturbation have been calculated numerically for the helium and beryllium atoms and the hydrogen molecule without introducing any density-functional approximations. In comparison with the RS-based perturbation theory, the present GL-based perturbation theory gives much more accurate excitation energies for Rydberg states but similar excitation energies for valence states.
\end{abstract}


\section{Introduction}

Within the framework of density-functional theory (DFT), the calculation of
molecular excitation energies is nowadays mostly performed using linear-response
time-dependent density-functional theory (TDDFT) (see, e.g.,
Refs.\,\onlinecite{Casida2009, Casida2012}) within the adiabatic local or
semi-local approximations.  Despite its many successes, linear-response TDDFT
within these approximations suffers from serious limitations, especially for
describing systems with static (or strong) correlation~\cite{Gritsenko2000},
double or multiple excitations~\cite{Maitra2004}, and Rydberg and
charge-transfer excitations~\cite{Casida1998,Dreuw2003}. These deficiencies have
been attributed to the locality of the approximated exchange--correlation
potential and kernel either in space (local and semi-local approximations) or in
time (adiabatic approximation). While the former is directly linked to functional
development in time-independent DFT, the latter is a problem specific to the
time-dependent formulation. However, time dependence is in principle not
required to describe excited states since by the Hohenberg--Kohn
theorem~\cite{Hohenberg1964} the time-independent ground-state
density contains all the information about the system including information
about its excited states.

Over the years, several time-independent DFT approaches for calculating
excitation energies have emerged and are still being developed: ensemble
DFT~\cite{Theophilou1979, Gross1988a, PasGidPer-PRA-13, Franck2013, Yang2014,
  Pribram-Jones2014, SenKneJenFro-PRA-15, SenHedAlaKneFro-MP-16,AlaKneFro-PRA-16, Filatov2016,AlaDeuKneFro-JCP-17}, state-specific self-consistent DFT and related
methods~\cite{Gunnarsson1976, Ziegler1977, Barth1979, Nag-IJQC-98, Gorling1999,
  Levy1999, Gor-PRL-00, NagLev-PRA-01, FerAss-JCP-02, Nag-IJQC-04, Harbola2004,
  VitDelGor-JCP-05, Gorling2005, SamHar-JPB-06, GluLev-JCP-07, AyeLev-PRA-09,
  KowYosVoo-JCP-11, AyeLevNag-PRA-12, Harbola2012, KryZie-JCTC-13,
  EvaShuTul-JPCA-13, Ziegler2016}, hybrid DFT/configuration
interaction (CI) methods~\cite{Gri-CPL-96, GriWal-JCP-99, BecStaBurBla-CP-08, KadVoo-JCP-10,
  Lyskov2016} and perturbation theory starting from the non-interacting Kohn--Sham (KS) Hamiltonian~\cite{GorLev-PRB-93,
  Gorling1995, Gorling1996, Filippi1997}. In a previous work~\cite{Rebolini2015}, we have explored further 
 this density-functional perturbation-theory approach with one key modification: As a zeroth-order Hamiltonian, instead of using the
  non-interacting KS Hamiltonian, we use a {\it partially interacting
    Hamiltonian} incorporating the {\it long-range} part only of the Coulomb
  electron--electron interaction, corresponding to an intermediate point along a
  range-separated adiabatic connection~\cite{Savin1996, Yan-JCP-98, Pollet2003,
    Savin2003, Toulouse2004, TeaCorHel-JCP-10b, Rebolini2014}. The partially interacting
  zeroth-order Hamiltonian is of course closer to the exact Hamiltonian than is
  the non-interacting KS Hamiltonian, thereby putting less demand on the
  perturbation theory.

In this previous work~\cite{Rebolini2015}, a Rayleigh--Schr\"odinger (RS)-based perturbation theory was tested on a few atoms and molecules and it
was found that the first-order excitation energies are not overall improved in comparison with the zeroth-order excitation energies. 
This finding was rationalized by the fact that this perturbation theory does not keep the ground-state density constant at each order. 
In the present work, we explore an alternative approach, based on Görling--Levy (GL) perturbation theory along a range-separated adiabatic connection, which keeps the ground-state density constant at each order and is expected to give more accurate excitation energies.

The paper is organized as follows. The main equations of our GL-based range-separated perturbation theory are given in Section~\ref{sec:rsdft}.
The computational details for the calculations carried out, involving no other approximations than the use of a finite basis, can be found in Section~\ref{sec:computational}. The results obtained for the He and Be atoms and for the H$_2$ molecule are discussed in Section~\ref{sec:results}. Finally, Section~\ref{sec:conclusion} contains our conclusions.

\section{Theoretical background}
\label{sec:rsdft}

This section consists of two parts. We first review range-separated DFT for ground states in Subsection\;\ref{sec:theory1} 
and then GL-based perturbation theory for excited states in Subsection\;\ref{sec:theory2}.

\subsection{Range-separated ground-state density-functional theory}
\label{sec:theory1}
In range-separated DFT (see, e.g., Ref.\,\onlinecite{Toulouse2004}), the
electron--electron interaction is partitioned into long-range and short-range
contributions by means of the error function and of a range-separation parameter
$\mu$ which controls the range of the separation.  The long-range (lr)
interaction operator is defined as
\begin{eqnarray}
  \hat{W}_{\ee}^{\lr,\mu} \!=\! \frac{1}{2} \iint
  w_{\ee}^{\lr,\mu}(r_{12}) \hat{n}_2(\mr_1,\mr_2) \mathrm d\mr_1
   \mathrm d\mr_2,
\end{eqnarray}
and is written in terms of the error-function interaction
$w_{\ee}^{\lr,\mu}(r)\!  =\!\erf(\mu r)/r$ and the pair-density
operator $\hat{n}_2(\mr_1,\mr_2)$, where $\mr$ refers to the electron
coordinates.
The exact ground-state energy of an $N$-electron system is then expressed as 
\begin{equation}
  E_0 = \min_{\Psi} \Bigl\{ \bra{\Psi} \hat{T} + \hat{V}_\text{ne} +
  \hat{W}_{\ee}^{\lr,\mu} \ket{\Psi} +
  \bar{E}_{\Hxc}^{\sr,\mu}[n_{\Psi}]\Bigl\},
\label{EminPsi}
\end{equation}
where the minimization is performed over normalized multi-determinantal wave
functions $\Psi$. In Eq.\,\eqref{EminPsi}, we have introduced the kinetic-energy
operator $\hat{T}$, the nuclear--electron attraction operator $\hat{V}_\text{ne} = \int
v_\text{ne}(\mr) \hat{n}(\mr) \mathrm d\mr$ written in terms of the
density operator $\hat{n}(\mr)$, and the complement short-range (sr)
Hartree--exchange--correlation density functional
$\bar{E}_{\Hxc}^{\sr,\mu}[n_\Psi]$ evaluated at the density of $\Psi$,
$n_\Psi(\mr) = \bra{\Psi} \hat{n}(\mr) \ket{\Psi}$.  The minimizing wave
function $\Psi_0^{\mu}$ in Eq.\,(\ref{EminPsi}) is the ground-state wave function
of the following eigenvalue problem
\begin{equation}
  \hat{H}^{\lr,\mu} | \Psi_0^{\mu} \rangle = {\cal E}_0^{\mu} |
  \Psi_0^{\mu} \rangle,
  \label{eq:eig}
\end{equation}
associated with the energy ${\cal E}_0^{\mu}$.  In Eq.\,\eqref{eq:eig}, we have
introduced the partially interacting Hamiltonian
\begin{equation}
  \hat{H}^{\lr,\mu} = \hat{T} + \hat{V}_\text{ne} +
  \hat{W}^{\lr,\mu}_{\ee} + \hat{\bar{V}}_{\Hxc}^{\sr,\mu},
  \label{Hmu}
\end{equation}
which contains the operator
\begin{equation}
  \hat{\bar{V}}_{\Hxc}^{\sr,\mu}= \int \bar{v}_{\Hxc}^{\sr,\mu}(\mr) \hat{n}(\mr) \mathrm d\mr,
\end{equation}
with the short-range Hartree--exchange--correlation potential $\bar{v}_{\Hxc}^{\sr,\mu}(\mr) = \delta \bar{E}_{\Hxc}^{\sr,\mu}[n_0]/\delta n(\mr)$ evaluated at the exact ground-state density $n_0(\mr)$. This is the potential that keeps the ground-state density constant for all $\mu$, i.e. $\bra{\Psi_0^{\mu}} \hat{n}(\mr) \ket{\Psi_0^{\mu}} = n_0(\mr)$. In this paper, contrary to what was sometimes done in previous papers on range-separated DFT, we use an overline in the notation for the short-range Hartree--exchange--correlation energy $\bar{E}_{\Hxc}^{\sr,\mu}[n]$ and its associated potential operator $\hat{\bar{V}}_{\Hxc}^{\sr,\mu}$ to emphasize that these quantities are defined as \textit{complements} to their long-range analogues, i.e. they include a mixed long-range/short-range correlation contribution~\cite{Tou-THESIS-05,TouSav-JMS-06}.

At $\mu=0$, the Hamiltonian $\hat{H}^{\lr,\mu}$ in
Eq.\,(\ref{Hmu}) reduces to the standard non-interacting KS Hamiltonian,
whereas for $\mu\to\infty$ it reduces to
the physical Hamiltonian.  The
Hamiltonian $\hat{H}^{\lr,\mu}$ therefore defines a range-separated adiabatic
connection, linking the KS and the physical systems when varying $\mu$, keeping
the ground-state density constant.

\subsection{Excited states from GL-based perturbation theory}
\label{sec:theory2}

For a given value of $\mu$, the excited-state wave functions and energies of the
long-range interacting Hamiltonian
\begin{equation}
  \hat{H}^{\lr,\mu} | \Psi_k^{\mu} \rangle = {\cal E}_k^{\mu} |
  \Psi_k^{\mu} \rangle,
\label{zerothordereq}
\end{equation}
can be used as zeroth-order approximations to the physical excited-state wave functions
and energies. They can then be improved upon by setting up perturbation theories with 
$\hat{H}^{\lr,\mu}$ as the zeroth-order
Hamiltonian.  As shown in Ref.\,\onlinecite{Rebolini2015}, an RS-based 
perturbation theory in which the ground-state density is not kept constant gives first-order excitation energies 
that overall do not improve upon the zeroth-order excitation energies. Here we instead use a GL-based perturbation theory, 
with the ground-state density kept constant.

To formulate such a GL-based perturbation theory, we define the following Hamiltonian
depending on a coupling constant $\la$
\begin{equation}
  \hat{H}^{\mu,\la} =
  \hat{T} + \hat{V}_\text{ne} + \hat{W}_{\ee}^{\lr,\mu} +
  \la \hat{W}_{\ee}^{\sr,\mu} + \hat{\bar{V}}_{\Hxc}^{\sr,\mu,\la},
  \label{eq:Hml}
\end{equation}
which contains the operator
\begin{equation}
  \hat{\bar{V}}_{\Hxc}^{\sr,\mu,\la} = \int \bar{v}_{\Hxc}^{\sr,\mu,\la}(\mr)
  \hat{n}(\mr) \, \mathrm d\mr,
\end{equation}
where $\bar{v}_{\Hxc}^{\sr,\mu,\la}(\mr)$ is the $\la$-dependent short-range Hartree--exchange--correlation potential that keeps the
ground-state density constant for all $\mu$ and all $\la$ -- that is, $\bra{\Psi_0^{\mu,\la}} \hat{n}(\mr) \ket{\Psi_0^{\mu,\la}} = n_0(\mr)$ where $\Psi_0^{\mu,\la}$ is the ground-state wave function of the Hamiltonian $\hat{H}^{\mu,\la}$ in Eq.\,\eqref{eq:Hml}. The Hamiltonian $\hat{H}^{\mu,\la}$ thus sets up a double adiabatic connection with a constant ground-state density.
To clearly separate the linear and non-linear dependence in $\la$, we then rewrite Eq.\,\eqref{eq:Hml} as
\begin{equation}
  \hat{H}^{\mu,\la}  =  \hat{T} + \hat{V}_\text{ne} + \hat{W}_{\ee}^{\lr,\mu} +
  \hat{\bar{V}}_{\Hxc}^{\sr,\mu} + \la \hat{W}_{\ee}^{\sr,\mu} +
  \hat{\bar{V}}_{\Hxc}^{\sr,\mu,\la} - \hat{\bar{V}}_{\Hxc}^{\sr,\mu},
  \label{Hlmu2}
\end{equation}
and we define the potential operator
\begin{equation}
  \hat{V}_{\Hxc}^{\sr,\mu,\la} = \hat{\bar{V}}_{\Hxc}^{\sr,\mu} -
  \hat{\bar{V}}_{\Hxc}^{\sr,\mu,\la},
\end{equation}
which we choose to denote without an overline because it is a ``double complement'' in the sense that it is the complement to the complement potential $\hat{\bar{V}}_{\Hxc}^{\sr,\mu,\la}$ with respect to the potential $\hat{\bar{V}}_{\Hxc}^{\sr,\mu}$. This potential can be decomposed into a linear contribution with respect to $\la$ and a term containing all higher-order terms (see Ref.\,\onlinecite{Rebolini2015})
\begin{equation}
  \hat{V}_{\Hxc}^{\sr,\mu,\la} = \la \hat{V}_{\H \x,\md}^{\sr,\mu} +
  \hat{V}_{\tc,\md}^{\sr,\mu,\la},
\end{equation}
where the potential $\hat{V}_{\H \x,\md}^{\sr,\mu}$ is the short-range
``multi-determinantal (md) Hartree--exchange'' potential operator, while
$\hat{V}_{\tc,\md}^{\sr,\mu,\la}$ is the $\la$-dependent short-range ``multi-determinantal
correlation'' potential operator (see Ref.~\cite{TouGorSav-TCA-05}). For non-degenerate ground states of the long-range interacting Hamiltonian in Eq.~(\ref{Hmu}), the expansion of $\hat{V}_{\tc,\md}^{\sr,\mu,\la}$ in
$\la$ around $\la=0$ begins at second order:
\begin{equation}
  \hat{V}_{\tc,\md}^{\sr,\mu,\la} = \la^2 \,
  \hat{V}_{\tc,\md}^{\sr,\mu,(2)} + \la^3 \,
  \hat{V}_{\tc,\md}^{\sr,\mu,(3)} + \cdots.
\end{equation}
The partially interacting Hamiltonian can then be rewritten as 
\begin{align}
\hat{H}^{\mu,\la}  & =  \hat{H}^{\lr,\mu} + \la \hat{W}_{\ee}^{\sr,\mu} -  \hat{V}_{\Hxc}^{\sr,\mu,\la}
\nonumber\\
                   & =  \hat{H}^{\lr,\mu} + \la \hat{W}^{\sr,\mu} -
\hat{V}_{\tc,\md}^{\sr,\mu,\la} \nonumber\\
& =   \hat{H}^{\lr,\mu} + \la \hat{W}^{\sr,\mu} - \la^2 \,
  \hat{V}_{\tc,\md}^{\sr,\mu,(2)} - \la^3 \,
  \hat{V}_{\tc,\md}^{\sr,\mu,(3)} + \cdots,
\end{align}
where the unperturbed Hamiltonian is the partially interacting Hamiltonian
defined in Eq.\,\eqref{Hmu} and $\hat{W}^{\sr,\mu} = \hat{W}_{\ee}^{\sr,\mu} -
\hat{V}_{\H \x,\md}^{\sr,\mu}$ is the perturbation operator.
More details on this GL-based perturbation theory can be found in the Appendix of Ref.\,\onlinecite{Rebolini2015}.

The excited-state wave functions and energies of the perturbed Hamiltonian are
thus expanded with respect to $\la$ as
\begin{align}
  \ket{\Psi_k^{\mu,\la}} &= \ket{\Psi_k^{\mu}} + \la \ket{\Psi_k^{\mu,(1)}}
        + \la^2 \ket{\Psi_k^{\mu,(2)}}  + \la^3 \ket{\Psi_k^{\mu,(3)}} + \cdots \\
  E_k^{\mu,\la} &= \mathcal{E}_k^{\mu} + \la E_k^{\mu,(1)} + \la^2
  E_k^{\mu,(2)} + \la^3  E_k^{\mu,(3)} + \cdots
\end{align}
where $\Psi_k^{\mu,(n)}$ and $E_k^{\mu,(n)}$ are the $n$-th order corrections to the
wave function and energy of the $k$-th state. In particular, the first-order energy correction is
\begin{eqnarray}
E_k^{\mu,(1)} = \bra{\Psi_k^{\mu}} \hat{W}^{\sr,\mu} \ket{\Psi_k^{\mu}},
\end{eqnarray}
and the corresponding zeroth+first order energy is
\begin{eqnarray}
E_k^{\mu,(0+1)} = \mathcal{E}_k^{\mu} + E_k^{\mu,(1)} = \bra{\Psi_k^{\mu}} \hat{H} + \hat{\bar{V}}_{\tc,\md}^{\sr,\mu} \ket{\Psi_k^{\mu}},
\label{Ekmu01}
\end{eqnarray}
where $\hat{\bar{V}}_{\tc,\md}^{\sr,\mu} = \hat{\bar{V}}_{\Hxc}^{\sr,\mu} -  \hat{{V}}_{\Hx,\md}^{\sr,\mu} = \hat{V}_{\tc,\md}^{\sr,\mu,\la=1}$ is the short-range multi-determinantal correlation potential operator. This operator can be expressed as $\hat{\bar{V}}_{\tc,\md}^{\sr,\mu} = \int \bar{v}_{\tc,\md}^{\sr,\mu}(\mr) \hat{n}(\mr) \text{d}\mr$ where $\bar{v}_{\tc,\md}^{\sr,\mu}(\mr)=\delta \bar{E}_{\tc,\md}^{\sr,\mu}[n_0]/\delta n(\mr)$ is the functional derivative of the short-range multi-determinantal correlation energy functional $\bar{E}_{\tc,\md}^{\sr,\mu}[n]$ introduced in Refs.~\cite{Tou-THESIS-05,TouGorSav-TCA-05} (see also Refs.~\cite{StoTeaTouHelFro-JCP-13,CorStoJenFro-PRA-13,CorFro-IJQC-14}), evaluated at the exact ground-state density $n_0$.
Equation~(\ref{Ekmu01}) contains two contributions; the zeroth+first order energy in standard RS perturbation theory, 
\begin{eqnarray}
E_{k,\text{RS}}^{\mu,(0+1)} =\bra{\Psi_k^{\mu}} \hat{H} \ket{\Psi_k^{\mu}},
\end{eqnarray}
and an additional term, 
\begin{eqnarray}
\bra{\Psi_k^{\mu}} \hat{\bar{V}}_{\tc,\md}^{\sr,\mu} \ket{\Psi_k^{\mu}}=\int \bar{v}_{\tc,\md}^{\sr,\mu}(\mr) n_{\Psi_k^{\mu}}(\mr) \text{d}\mr,
\label{addGL}
\end{eqnarray}
which is only present in GL-based perturbation theory.
For $\mu=0$, the zeroth-order wave functions are KS single determinants, $\Psi_k^{\mu=0}=\Phi_k$, and the short-range multi-determinantal correlation potential reduces to the standard KS potential, $\bar{v}_{\tc,\md}^{\sr,\mu=0}(\mr) = v_{\tc}(\mr)$, and we thus recover the standard GL perturbation theory for excitation energies~\cite{Gorling1996, Filippi1997}.

A local-density approximation has been developed for the short-range multideterminant correlation energy functional $\bar{E}_{\tc,\md}^{\sr,\mu}[n]$ (and thus for its associated potential $\bar{v}_{\tc,\md}^{\sr,\mu}(\mr)$)~\cite{TouGorSav-TCA-05,PazMorGorBac-PRB-06}. In the present study, we will test the GL-based first-order perturbation theory without introducing any density-functional approximations, providing benchmark data for future approximations.

\section{Computational details}
\label{sec:computational}

Calculations were performed for the He and Be atoms and the H$_2$ molecule with
a development version of the DALTON program~\cite{dalpaper,DaltonWeb}; see
Refs.\,\onlinecite{TeaCorHel-JCP-09,TeaCorHel-JCP-10,TeaCorHel-JCP-10b}.
Following the same procedure as in Ref.\,\onlinecite{Rebolini2014}, a full CI
(FCI) calculation was first carried out to obtain the exact ground-state density
in the chosen basis set. Next, for several values of $\mu$ and $\la$, a Lieb optimization~\cite{ColSav-JCP-99,WuYan-JCP-03,TeaCorHel-JCP-09} was carried out to obtain the short-range
potential $\bar{v}^{\sr,\mu, \la}(\mr) = v_\text{ne}(\mr) + \bar{v}^{\sr,\mu, \la}_\Hxc(\mr)$ needed in Eq.\,\eqref{eq:Hml} to reproduce 
this FCI ground-state density with the partial electron--electron interaction $\w^{\lr,\mu}(r_{12}) +
\la\w^{\sr,\mu}(r_{12}) $. Then, an FCI calculation was performed with the
partially interacting Hamiltonian constructed from $\w^{\lr,\mu}(r_{12}) + \la
\w^{\sr,\mu}(r_{12})$ and $\bar{v}^{\sr,\mu, \la}(\mr)$ to obtain, for a few states, the
energies $E_k^{\mu,\la}$ and wave functions $\Psi_k^{\mu,\la}$, for each values of $\mu$ and $\la$.

For each system, each excited state, and each value of $\mu$, a third-degree polynomial fit in $\la$ was
performed on the excitation energy $\Delta E_k^{\mu,\la} = E_k^{\mu,\la} - E_0^{\mu,\la}$ for values of $\la$ ranging from 0 to 0.3 and the first-order correction in the GL-based perturbation theory was obtained as the first-order derivative with respect to $\la$. The zeroth-order excitation energies and the first-order excitation energies in the RS-based perturbation theory
were already available from Refs.\,\onlinecite{Rebolini2014} and~\onlinecite{Rebolini2015}, respectively.
Because of issues with loss of numerical precision, we refrain from extracting second- and higher-order derivatives with respect to $\la$, which could be used to test higher-order perturbation theories or extrapolation schemes~\cite{RebTouTeaHelSav-PRA-15}.

The basis sets used were uncontracted t-aug-cc-pV5Z for He, uncontracted
d-aug-cc-pVDZ for Be, and uncontracted d-aug-cc-pVTZ for H$_2$.

\section{Results and discussion}
\label{sec:results}

\subsection{First Rydberg excitation energies of the helium atom}

\begin{figure}
  \includegraphics[scale=1]{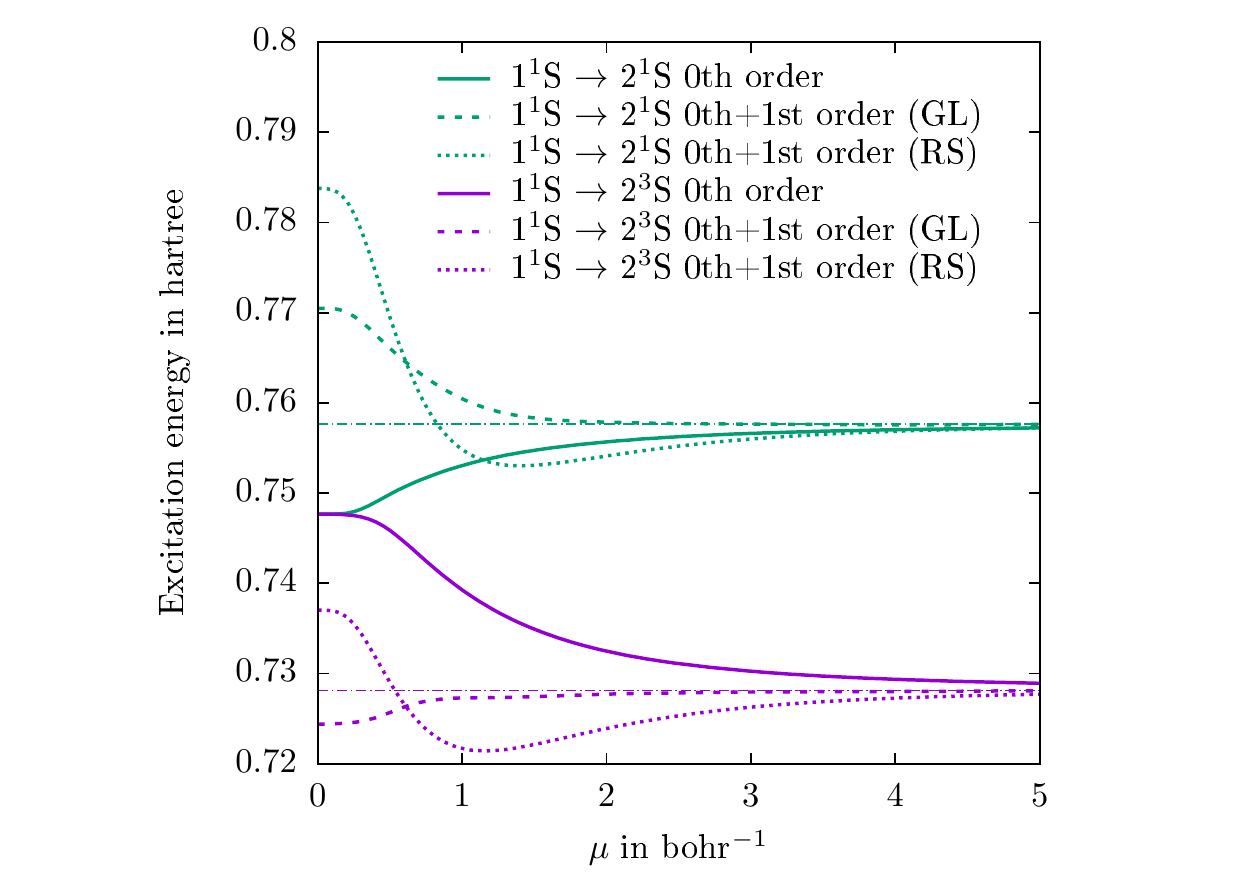}
  \includegraphics[scale=1]{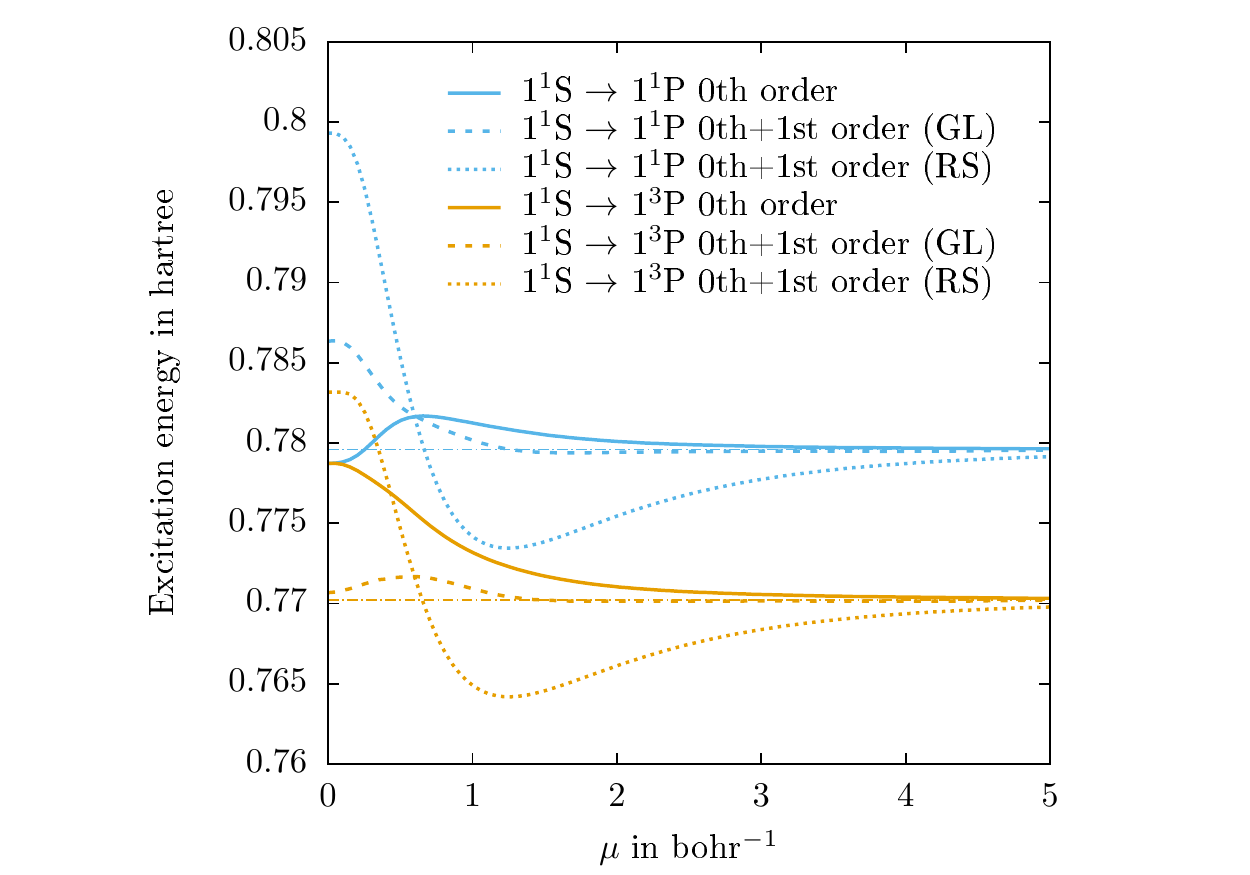} 
  \caption{Zeroth-order excitation energies $\Delta {\cal E}_k^{\mu}={\cal E}_k^{\mu}-{\cal E}_0^{\mu}$ (plain line) and zeroth+first order excitation energies $\Delta { E}_k^{\mu,(0+1)}=E_k^{\mu,(0+1)} - E_0^{\mu,(0+1)}$ (dashed line) in the GL-based perturbation theory for the helium atom as a function of $\mu$. For comparison, the zeroth+first order excitation energies $\Delta {E}_{k,\text{RS}}^{\mu,(0+1)}=E_{k,\text{RS}}^{\mu,(0+1)} - E_{0,\text{RS}}^{\mu,(0+1)}$ (dotted line) in the RS-based perturbation theory of Ref.\,\onlinecite{Rebolini2015} are also shown. The excitation energies of the physical system $\Delta E_k=E_k-E_0$ are plotted as thin horizontal dot-dashed lines.
    \label{fig:he}
  }
\end{figure}

The first singlet and triplet excitation energies of the helium atom correct to
zeroth and first orders along the range-separated adiabatic
connection in the GL- and RS-based perturbation theories are shown in Figure\,\ref{fig:he}.

In the KS limit, at $\mu=0$, the zeroth-order singlet and triplet excitation energies are degenerate. When increasing $\mu$, this degeneracy is lifted and the zeroth-order singlet and triplet excitation energies eventually converge to their physical excitation energies for $\mu \to \infty$. With the introduction of the perturbation, the singlet/triplet degeneracy is lifted already at $\mu=0$. 
As found in Ref.\,\onlinecite{Rebolini2015}, the RS-based first-order perturbative correction globally deteriorates the zeroth-order excitation energies, leading to large errors that do not decrease monotonically with $\mu$.

Interestingly, the zeroth+first order excitation energies obtained from the GL-based perturbation theory have smaller errors 
than those obtained by the RS-based perturbation theory. At $\mu=0$, the contributions to the excitation energies in the GL-based first-order perturbation theory coming from the additional term in Eq.~(\ref{addGL}) are equal to $\int v_{\tc}(\mr) [n_{\text{2s}}(\mr) - n_{\text{1s}}(\mr)]\text{d}\mr$ for the $1 {^1}\text{S} \to 2 {^1}\text{S}$ and $1 {^1}\text{S} \to 2 {^3}\text{S}$ transitions and $\int v_{\tc}(\mr) [n_{\text{2p}}(\mr) - n_{\text{1s}}(\mr)]\text{d}\mr$ for the $1 {^1}\text{S} \to 1 {^1}\text{P}$ and $1 {^1}\text{S} \to 1 {^3}\text{P}$ transitions, respectively, where $n_{k}(\mr)$ is the density of the KS orbital $k$. At the scale of the plot, these corrections to the RS-based first-order perturbative excitation energies are significant (more than $-0.01$ eV) and strongly reduce the errors. As $\mu$ increases, the corrections coming from the additional term in Eq.~(\ref{addGL}) change sign but are still efficient to reduce the errors. For sufficiently large $\mu$, in comparison to the zeroth-order excitation energies and to the zeroth+first-order RS excitation energies, the zeroth+first-order GL excitation energies systematically converge faster with respect to the range-separation parameter to the physical excitation energies: a 1 millihartree accuracy is reached for $\mu$ larger than 1.2 bohr$^{-1}$, while values of 3 to 4 bohr$^{-1}$ are necessary for the zeroth-order curves. However, near $\mu=0$, the first-order GL correction does not always improve the zeroth-order excitation energy (see the $1 ^1\text{S} \to 1 ^1\text{P}$ transition). 

These comparisons show that the GL-based perturbation theory (which keeps the ground-state density constant, ensuring a correct ionization energy) 
is a much better strategy for calculating Rydberg excitation energies than the RS-based perturbation theory.
This was expected since these Rydberg excitation energies are close to the ionization threshold.

\subsection{Valence excitation energies of the beryllium atom}

\begin{figure}[h!]
  \includegraphics[scale=1]{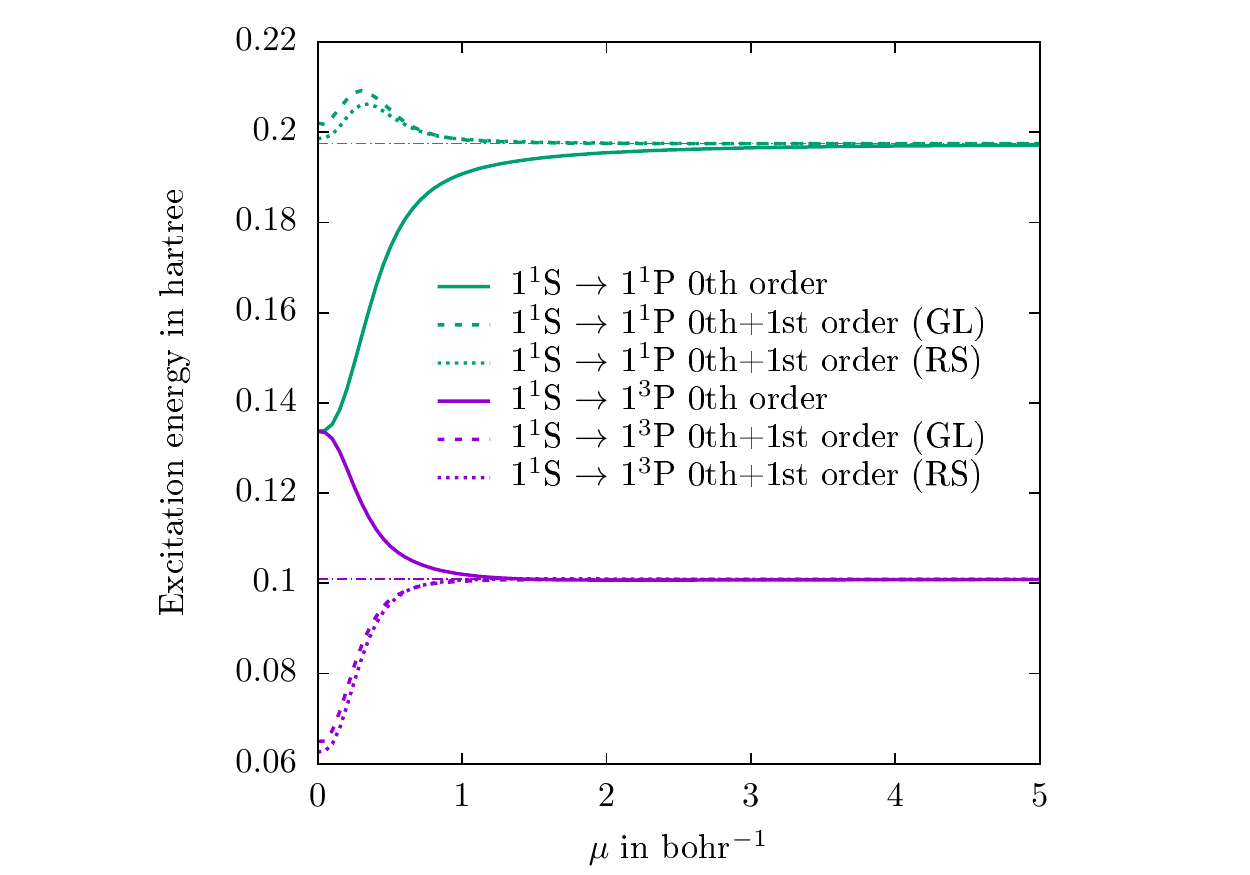}
  \caption{Zeroth-order excitation energies $\Delta {\cal E}_k^{\mu}={\cal E}_k^{\mu}-{\cal E}_0^{\mu}$ (plain line) and zeroth+first order excitation energies $\Delta { E}_k^{\mu,(0+1)}=E_k^{\mu,(0+1)} - E_0^{\mu,(0+1)}$ (dashed line) in the GL-based perturbation theory for the beryllium atom as a function of $\mu$. For comparison, the zeroth+first order excitation energies $\Delta {E}_{k,\text{RS}}^{\mu,(0+1)}=E_{k,\text{RS}}^{\mu,(0+1)} - E_{0,\text{RS}}^{\mu,(0+1)}$ (dotted line) in the RS-based perturbation theory of Ref.\,\onlinecite{Rebolini2015} are also shown. The excitation energies of the physical system $\Delta E_k=E_k-E_0$ are plotted as thin horizontal dot-dashed lines.
    \label{fig:be}
  }
\end{figure}

The valence excitation energies of the beryllium atom correct to zeroth and first
orders in the GL-based and RS-based perturbation theories are plotted in Figure\,\ref{fig:be}. 
For these excitations, the two approaches give very similar first-order corrections. 
This behaviour can be rationalized from the fact that these valence excited states are far from the ionization threshold 
so that  imposing the correct ionization energy (by keeping the ground-state density constant)
has much less impact than for the Rydberg excitation energies. 
This can also be understood by looking at the expression of the difference between the zeroth+first-order GL and RS excitation energies coming from the additional term in Eq.~(\ref{addGL}) which is, for $\mu=0$, equal to $\int v_{\tc}(\mr) [n_{\text{2p}}(\mr) - n_{\text{2s}}(\mr)]\text{d}\mr$ for the $1 {^1}\text{S} \to 1 {^1}\text{P}$ and $1 {^1}\text{S} \to 1 {^3}\text{P}$ transitions. This quantity is necessarily small since the 2s and 2p orbitals are localized in the same region of space.

We note that the first-order correction systematically improves the singlet excitation energy but not the triplet excitation energy, which
is overestimated at zeroth order and underestimated by about the same amount at zeroth+first order for small $\mu$.

\begin{figure}[h!]
  \includegraphics[scale=1]{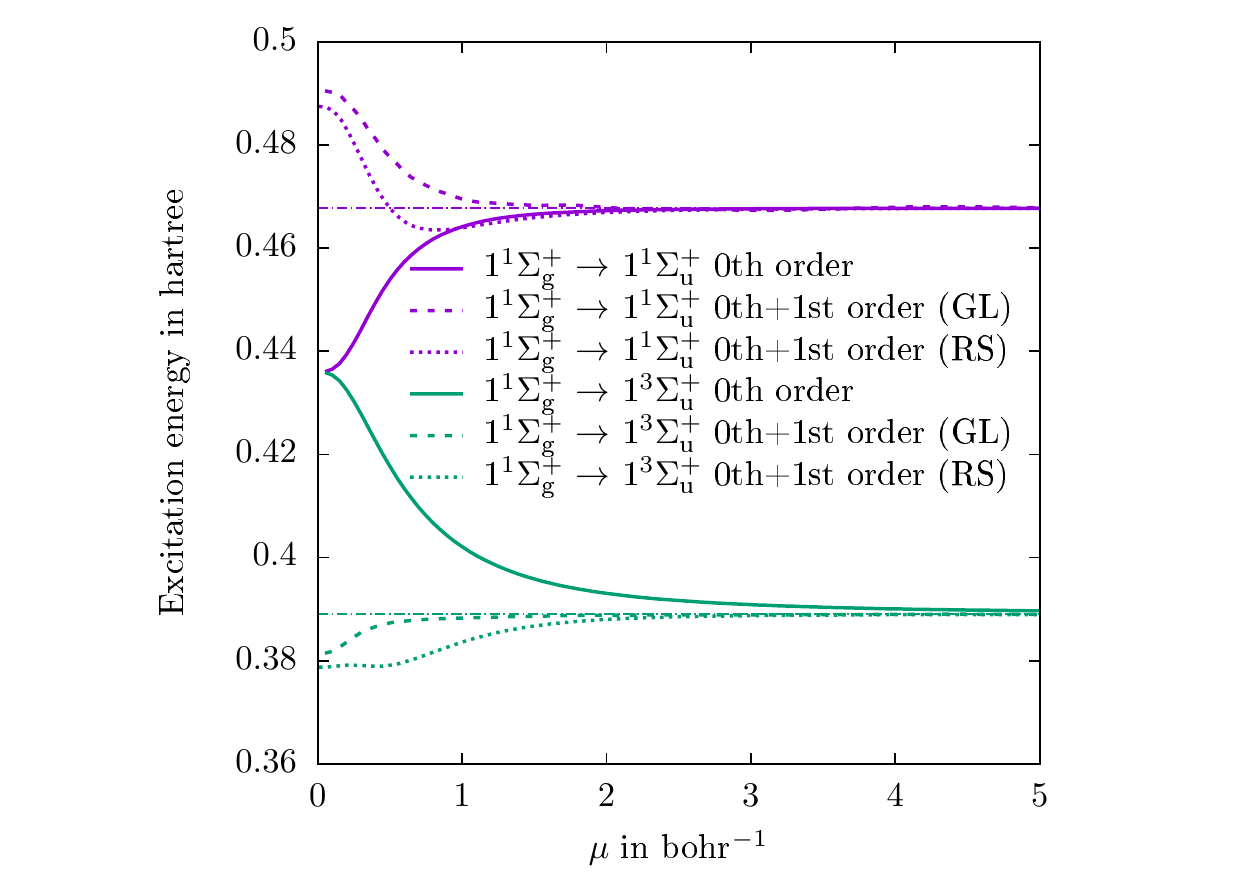}
  \includegraphics[scale=1]{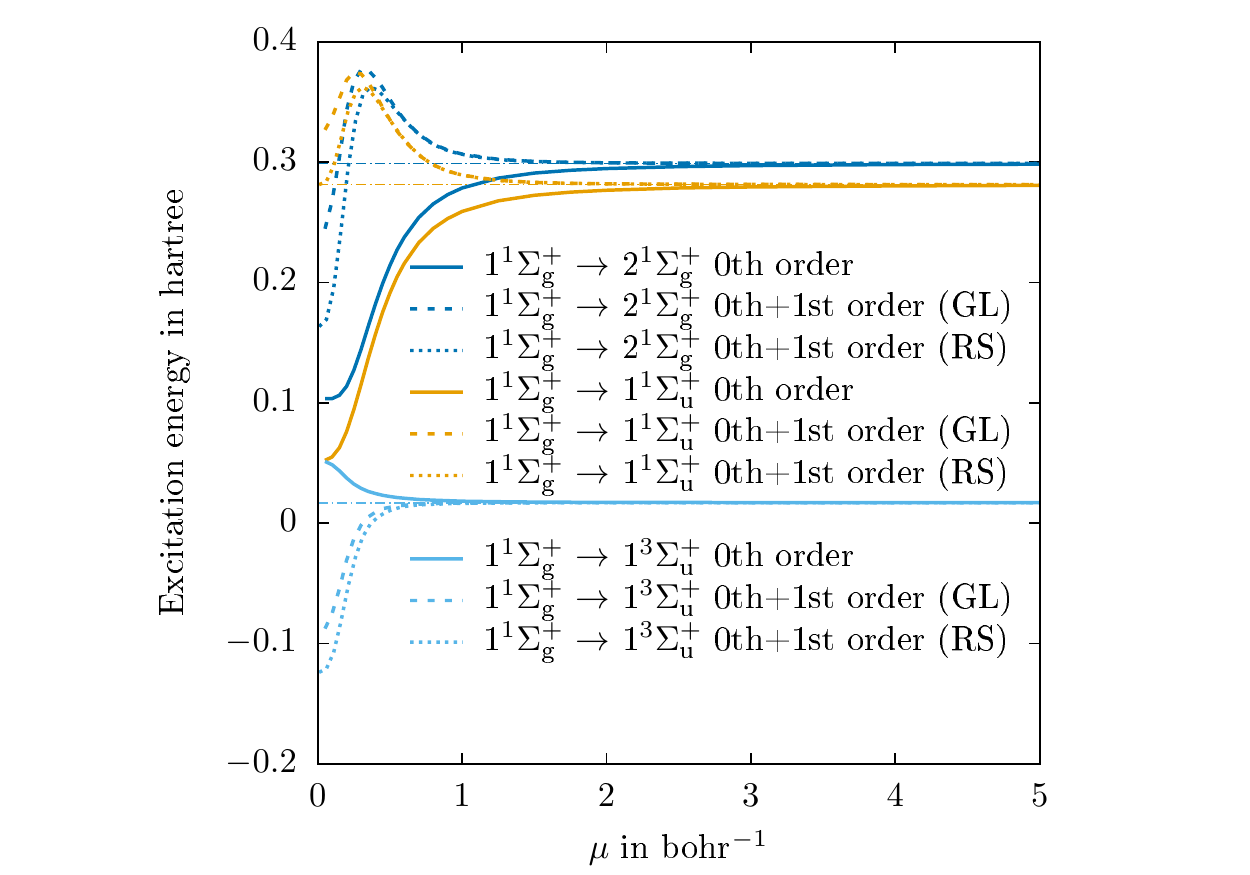}
  \caption{Zeroth-order excitation energies $\Delta {\cal E}_k^{\mu}={\cal E}_k^{\mu}-{\cal E}_0^{\mu}$ (plain line) and zeroth+first order excitation energies $\Delta { E}_k^{\mu,(0+1)}=E_k^{\mu,(0+1)} - E_0^{\mu,(0+1)}$ (dashed line) in the GL-based perturbation theory for the hydrogen molecule at the equilibrium distance (top) and three times the equilibrium distance (bottom) as a function of $\mu$. For comparison, the zeroth+first order excitation energies $\Delta {E}_{k,\text{RS}}^{\mu,(0+1)}=E_{k,\text{RS}}^{\mu,(0+1)} - E_{0,\text{RS}}^{\mu,(0+1)}$ (dotted line) in the RS-based perturbation theory of Ref.\,\onlinecite{Rebolini2015} are also shown. The excitation energies of the physical system $\Delta E_k=E_k-E_0$ are plotted as thin horizontal dot-dashed lines.
    \label{fig:h2}
    }
\end{figure}

\subsection{First valence excitation energies of the hydrogen molecule}
 
In Figure~\ref{fig:h2}, we have plotted the first valence excitation energies of the hydrogen molecule 
at its equilibrium internuclear distance $R_\text{eq}$ and at 3$R_\text{eq}$.

As for the valence excitation energies in beryllium, the first-order RS- and GL-based corrections are overall quite similar, but
with a discernible faster $\mu$-convergence of the zeroth+first-order GL excitation energies to the physical excitation energies at the equilibrium distance. Again, the fact that the difference between the zeroth+first-order GL and RS excitation energies is small can be understood from its expression at $\mu=0$ which is $\int v_{\tc}(\mr) [n_{1\sigma_\text{u}}(\mr) - n_{1\sigma_\text{g}}(\mr)]\text{d}\mr$ for the $1 ^1\Sigma_\text{g}^+ \to 1 ^1\Sigma_\text{u}^+$ and $1 ^1\Sigma_\text{g}^+ \to 1 ^3\Sigma_\text{u}^+$ transitions, and which involves the difference between two similar orbital densities. At the equilibrium distance, the GL-based first-order perturbation theory overshoots the correction to both the singlet and triplet zeroth-order excitation energies for small values of $\mu$, but nevertheless improves upon the zeroth-order correction for all $\mu$ values. 

When the bond is stretched, the first-order correction no longer systematically improves on the zeroth-order excitation energies for small $\mu$. The zeroth+first-order excitation energy of the first transition $1 ^1\Sigma_\text{g}^+ \to 1 ^3\Sigma_\text{u}^+$ becomes negative for small $\mu$ and the error with respect to the physical excitation energy is higher than in the zeroth-order case. The zeroth+first-order excitation energies for the two singlet states, $1 ^1\Sigma_\text{u}^+$ and $2 ^1\Sigma_\text{g}^+$,  are incorrectly ordered at small $\mu$ and do not monotonically converge to the physical energies, passing through a maximum at around $\mu\approx 0.5$ bohr$^{-1}$. At $\mu=0$, in comparison to the RS-based first-order perturbation theory, the GL-based first-order perturbation theory gives a better estimate of the excitation energy for the $2 ^1\Sigma_\text{g}^+$ state of double-excitation character, the additional term in the first-order GL-based correction being $2\int v_{\tc}(\mr) [n_{1\sigma_\text{u}}(\mr) - n_{1\sigma_\text{g}}(\mr)]\text{d}\mr$. However, in the dissociation limit, it is easy to show that both RS-based and GL-based first-order perturbation theories will incorrectly lead to a vanishing excitation energy at $\mu=0$ for this state. Only for values of $\mu$ greater than about 1 bohr$^{-1}$, both RS- and GL-based first-order perturbation theories provide accurate estimates of the excitation energies for the all states considered.

\section{Conclusion}
\label{sec:conclusion}

We have applied a GL-based perturbation theory along a range-separated adiabatic connection for the calculation of electronic excitation energies. 
Unlike the RS-based perturbation theory that we explored in a previous work, the GL-based perturbation theory keeps the ground-state density (and thus
the ionization energy) constant at each order. 
Excitation energies up to first order in the perturbation have been calculated numerically for the helium and beryllium atoms and the hydrogen molecule without introducing any density-functional approximations.

In comparison with the RS-based perturbation theory, the GL-based perturbation theory gives much more accurate excitation energies for the Rydberg states of the helium atom but similar excitation energies for the valence states of the beryllium atom and of the hydrogen molecule. This can be rationalized by observing that the Rydberg states are 
close to the ionization threshold and therefore sensitive to having the correct ionization energy. 

This first-order GL-based perturbation theory works reasonably well for calculating the first valence excitation energies of the hydrogen molecule at its equilibrium distance. However, results are less satisfactory for the valence excitation energies of the beryllium atom and stretched hydrogen molecule at small range-separation parameter. 
For such systems, with small HOMO-LUMO gaps, it may be necessary to go beyond single-reference first-order perturbation theory for
small range-separation parameters.

One possible extension of this work would be to test density-functional approximations for the short-range exchange–correlation potential and approximations for the wave-function part of the calculation. In particular, the present approach could be applied in practice by first performing a range-separated DFT ground-state calculation using a long-range multiconfiguration self-consistent-field (MCSCF) wave function and a short-range (semi-)local density-functional approximation in Eq.~(\ref{EminPsi}), as developed in Refs.~\cite{PedJen-JJJ-XX,FroTouJen-JCP-07,FroReaWahWahJen-JCP-09,HedTouJen-ARX-17}, then doing a CI calculation for the excited states in Eq.~(\ref{zerothordereq}), and evaluating the excitation energies via Eq.~(\ref{Ekmu01}) using a density-functional approximation for $\bar{v}_{\tc,\md}^{\sr,\mu}(\mr)$. We would then obtain a time-independent range-separated DFT method for calculating excitation energies, as an alternative to more usual linear-response time-dependent range-separated DFT approaches~\cite{RebSavTou-MP-13,FroKneJen-JCP-13,HedHeiKneFroJen-JCP-13,RebTou-JCP-16}. Finally, for nearly degenerate systems, it would also be interesting to explore the extension of this approach to range-separated ensemble DFT~\cite{SenKneJenFro-PRA-15,SenHedAlaKneFro-MP-16,AlaKneFro-PRA-16,AlaDeuKneFro-JCP-17} which would involve replacing the initial ground-state MCSCF calculation by a state-average MCSCF.

\section*{Acknowledgements}
This work was supported by the Norwegian Research Council through the
CoE Centre for Theoretical and Computational Chemistry (CTCC) Grant
No.\ 179568/V30 and through the European Research Council under
the European Union Seventh Framework Program through the Advanced
Grant ABACUS, ERC Grant Agreement No.\ 267683. A. M. T. gratefully acknowledges support from the Royal Society University Research Fellowship scheme and the Engineering and Physical Sciences Research Council (EPSRC), Grant No. EP/M029131/1.

\end{document}